\documentclass[paper,nofootinbib] {revtex4}

\usepackage{graphicx}
\usepackage{epsfig}
\usepackage{fancyhdr}
\usepackage{mathbbol}
\usepackage{pstricks}
\usepackage{color}
\usepackage{bbm}
\usepackage{multirow}
\usepackage{amsmath}
\def\mol{D^0\bar D^{*0}}

\def\be{\begin{equation}}
\def\ee{\end{equation}}
\def\bea{\begin{eqnarray}}
\def\eea{\end{eqnarray}}

\def\XX{X(3872)}
\def\ddsn{D^{0*}\bar{D}^0} 
 
\def\an#1#2#3{^{#1}\textrm{#2}_{#3}}     
\def\ev#1{\langle #1 \rangle}
\def\uP{\textrm P}
\def\uD{\textrm D}
\def\k|#1>{\ket{#1}}
\newcommand{\ket}[1]{|           {#1}           \rangle}
\def\Jp{J/\psi}
\def\b<#1|{\bra{#1}} 
\newcommand{\bra}[1]{\langle           {#1}           |}   
\def\bk<#1|#2>{\langle #1 | #2 \rangle}

\begin{document}

\title{The $2^{-+}$ assignment  for the $X(3872)$}

\author{T. J. Burns$^{\P}$,  F. Piccinini$^{\dag}$, A. D. Polosa$^\P$} 
\author{C. Sabelli$^{\ddag,\P}$}
\affiliation{
$^\P$INFN Roma, Piazzale A. Moro 2, Roma, I-00185, Italy\\
$^\dag$ INFN Pavia, Via A. Bassi 6, Pavia, I-27100, Italy\\
$^\ddag$Department of Physics, Universit\`a di Roma, `La Sapienza', Piazzale A. Moro 2, Roma, I-00185, Italy}

\begin{abstract}
Very recently the BaBar collaboration has put forward 
a claim that the $X(3872)$ is not a $1^{++}$ resonance, 
as most of the phenomenological work on the subject was 
relying on,  but rather a $2^{-+}$ one. We examine the 
consequences of this quantum number assignment for the 
solution of the $X(3872)$ puzzle. The molecular 
interpretation appears less likely, and the conventional 
charmonium interpretation should be reconsidered. 
There are several well-known difficulties with this 
interpretation, to which we add a new one: the production 
cross section at CDF is predicted to be much smaller than 
that observed. We also confirm, using a relativistic 
string model, the conclusion from potential models that 
the mass of the state is not consistent with expectations. 
In the tetraquark interpretation the $2^{-+}$ assignment 
implies a rich spectrum of partner states, although 
the $\XX$ may be among the few which are narrow enough to be observable.
\end{abstract}

\maketitle

\thispagestyle{fancy}

\section{Introduction}
In a very recent paper the BaBar collaboration claims~\cite{babarbn} that the quantum numbers of the $X(3872)$ are not $1^{++}$, as had generally been accepted, but $2^{-+}$.
Whereas an early conference paper by Belle~\cite{belleangular} favored $1^{++}$ quantum numbers, a later analysis by CDF with higher statistics~\cite{cdfangular} allowed both $1^{++}$ and $2^{-+}$. The new conclusion from BaBar is based on the distribution of the $3\pi$ mass in the $J/\psi\omega$ decay near threshold, which strongly favors a P-wave rather than S-wave decay. The earlier Belle result which favored $1^{++}$ arose from the same sort of analysis applied to the $2\pi$ mass distribution in $J/\psi\rho$.

If this quantum number assignment is confirmed, it would remove the main motivation for the molecular interpretation of the $X(3872)$, which assumes a loosely bound state of a $D^0$ and a $\bar D^{*0}$ in S-wave. It would also imply that the tetraquark hypothesis should be reexamined. Could it be, on the other hand, that the $X(3872)$ is a standard $1\an 1D2$ charmonium? Beyond the known problems of its isospin violation in $J/\psi\,\rho$ and $J/\psi\,\omega$ \cite{AbeEtAl05evidence}, its larger than expected $J/\psi\gamma$ and $\psi^{'}\gamma$ transitions \cite{AbeEtAl05evidence,AubertEtAl06search,JiaSangEtAl10is}, and its high mass \cite{Barnes:2003vb}, we would at least  understand why the $X(3872)$ is so narrow in the most prominent hadronic modes: they are close-to-threshold P-wave decays.

We shall show (Sec.~\ref{production}) that there is another reason why the $1\an 1D2$ assignment is not so straightforward. Relying on an earlier study for D-wave charmonium fragmentation \cite{chowise}, we calculate the expected prompt production cross section at CDF and find that it is considerably lower than the observed $X(3872)$ cross section; we also revisit the issue of the prompt production of the $1^{++}$ $\XX$ in the molecular interpretation. As a $1\an 1D2$ charmonium, the mass of the $\XX$ is known to disagree with potential model predictions. We investigate here a different approach, a relativistic string model in the heavy quark limit (Sec. \ref{stringmodel}). The model agrees remarkably well with the masses of established charmonia and bottomonia, but as in other approaches the $\XX$ is very difficult to accommodate as a $1\an 1D2$ state. We discuss possible decay modes that may help isolate the $1\an 1D2$ assignment (Sec. \ref{charmoniumdecays}). We also examine how to fit a $2^{-+}$ $X(3872)$ in a tetraquark model (Sec. \ref{4qsection}). The main drawback of this interpretation is the proliferation of predicted partner states, including many with masses close to the low-lying charmonia. We discuss the possibility (Sec. \ref{4qsectiondecays}) that the $\XX$ may be among the few which are narrow enough to be stable.

The news about the $X(3872)$ quantum numbers, {\it if confirmed}, will help discriminate among its possible interpretations.

\section{$X(3872)$ production at the Tevatron: molecule and charmonium}\label{production}
In a recent paper~\cite{bignamini} we proposed a method for estimating the prompt production cross section of $X(3872)$~\cite{bellex} 
at the Tevatron making the assumption that it is a loosely bound molecule of $D^0$ and  $\bar D^{*0}$, with a binding energy as small 
as ${\cal E}_0=-0.25\pm 0.40$~MeV. The motivation for this study is that, after the Belle discovery~\cite{bellex}, CDF and D0 confirmed the $X(3872)$ 
in proton-antiproton collisions~\cite{cdfx,d0x} and it  seems at least counter-intuitive that  such a loosely bound molecule could be 
produced promptly ({\it i.e.}, not from $B$ decay) in a high energy hadron collision environment. This was also one of the initial motivations to 
consider the possibility that the $X(3872)$ could be, instead of a molecule, a ``pointlike'' hadron resulting from  the binding of a diquark 
and an antidiquark~\cite{4q}, following the interpretation proposed by Jaffe and Wilczek~\cite{jw} of pentaquark baryons 
(antidiquark-antidiquark-quark). 

Indeed an analysis by CDF~\cite{cdfdata} allows one to distinguish the fraction of $X(3872)$ produced promptly from the one originated from $B$-decays.
The result of this analysis is
\be
\sigma(p\bar p\to X(3872))_{\rm prompt}\times \mathcal{B}(X\to J/\psi\pi^+\pi^-)= (3.1\pm0.7)~{\rm nb}. 
\ee
Following~\cite{braaten}, the experimental bounds on $ \mathcal{B}(X\to J/\psi\pi^+\pi^-)$ can be used to estimate a range for the prompt production cross section,
\begin{equation}
\sigma(p\bar p\to X(3872))_{\rm prompt}\simeq 30\div 70~{\rm nb} .
\label{exptvalue}
\end{equation}
In Ref.~\cite{bignamini} we estimated the maximum value of the theoretical prompt production 
cross section in the hypothesis that the $X(3872)$ is a $1^{++}$ molecular state with a tiny binding energy ${\cal E}_0$. Using standard MC tools, like HERWIG 
and PYTHIA, we computed the differential distribution of the prompt production cross section
with respect to $k$, the modulus of the spatial component of the relative three momentum in the center of mass of the mesons constituting the molecular $X(3872)$.
This is a relevant variable for the system since we expect the two mesons to be almost collinear, 
otherwise the molecule would break down immediately (on a time scale $\tau\sim1/k$).
In the MC simulation we have taken into account the kinematical cuts required for a comparison with CDF data.
Integrating the differential distribution up to $k\lesssim50~{\rm MeV}$ (an upper limit derived by the binding energy scale $k\approx\sqrt{2 m {\cal E}_0}$) 
we find that
\be
\sigma(p\bar p\to X(3872))_{{\rm prompt}}\; \lesssim 0.1~{\rm nb},
\ee
about {\it 300 times smaller} than the measured value.
This is a serious challenge to the molecular interpretation of the $X(3872)$. 

On the other hand, in Ref.~\cite{braaten} it was argued 
that final state interactions (FSI) in the
$\mol$ system require two corrections to our previous calculation: 
$(i)$ $k$ should range up to $\Lambda\sim 300$~MeV; $(ii)$ a correction factor to the cross section we compute should be included. This allows a spectacular reconcilement of the molecular picture with data, and Ref.~\cite{braaten} fully recovers the experimental prompt production cross section.

In Ref.~\cite{noiwats} we cast some doubts on the possibility that FSI can indeed play such a pivotal role. First, the Watson formulas~\cite{watson} used in~\cite{braaten} are valid for S-wave scattering, whereas a relative three-momentum $k$ of 
$300$~MeV indicates that higher partial waves should be taken into account.  Most importantly, we have verified in our MC simulations that, as the relative momentum $k$  in the center of mass of the molecule is taken to be up to $300$~MeV, then other hadrons (on average more than two) have a relative momentum $k<100$~MeV with respect to the $D$ or the $D^*$ 
constituting the molecule. The ``extra'' hadrons are not only pions.
On the other hand the Migdal-Watson theorem for FSI
requires that {\it only two} particles in the final state participate in the strong interactions causing them to rescatter. The extra hadrons involved in the process interfere in an unknown way with the mesons assumed to rescatter into an $X(3872)$. This is particularly true as one further enlarges the maximum value allowed for $k$, as required in~\cite{braaten}.

In a recent paper~\cite{braaten-last} it is argued that the latter problem can 
be overcome as the interaction between the $D$ mesons constituting the 
molecule is stronger than that between a $D^{(*)}$ and one of the additional 
hadrons. Only a qualitative explanation of this argument is given. 
Moreover, the use of the MC approach to estimate the production of loosely 
bound hadron molecules proposed in~\cite{bignamini} and then followed 
in~\cite{braaten} is criticized in \cite{braaten-last}. 
We observe that: $(i)$ 
The authors of~\cite{braaten-last} apply the method of~\cite{bignamini} 
to deuteron production and, playing with the $k$ parameter, they 
obtain disagreement with the empirical deuteron cross section 
by factors of ${\cal O}(1)$, even if the MC generator is not tuned 
on baryon pair production data. 
We stress here that in the case of the $X(3872)$ the MC cross section 
value is about $0.3\%$ of the observed one, after having tuned 
the MC generator on $D^0 D^{*-}$ data.
 $(ii)$ The deuteron is a system qualitatively different 
from the $DD^*$ molecule, whose components cannot have spin interactions 
since the $D$ is a spinless particle. 
In the case of  the deuteron  spin interactions play an important 
role in the determination of the binding: the spin $S=0$ deuterium 
is not formed. 
  
In our view the molecule picture is seriously challenged by the CDF cross section results. In any case if the $2^{-+}$ quantum numbers are confirmed then the motivation for the molecular interpretation, which assumes S-wave binding, no longer applies.
A $2^{-+}$ state formed out of
$\ddsn$ would require a relative P-wave, and it is unlikely that
$\pi$ exchange could bind such a state, given that even in S-wave it is
not clear that the attraction is sufficiently strong~\cite{Suzuki:2005ha}. Even if
such a state exists, there remains the further problem that unless
spin-dependent forces prevent the binding, one should expect
partner states with $0^{-+}$ and the $J^{PC}$-exotic $1^{-+}$
state, for which there is no experimental evidence. 
A P-wave $2^{-+}$ molecule would also imply the existence of a more
deeply bound S-wave $1^{++}$ molecule, which would be extremely narrow.
Alternatively, to form a $2^{-+}$ bound state in S-wave would require
$D_2D$ or $D_1D^*$, which not only implies an immense binding energy of
some 500 MeV, but one loses the appealing connection between the mass of
the $\XX$ and the $\ddsn$ threshold.

What about the prompt production rate if the $X(3872)$ is a $1 \an 1D2$ standard charmonium?
On general grounds one expects it to be small. The production cross section is proportional to a fragmentation function, which describes the probability that quarks and gluons hadronize into bound states. This function can be expressed as a perturbative expansion in the quark velocity $v$. For the production of a $1 \an 1D2$ state the function begins at order $O(v^7)$, and so one expects a smaller cross section than for 1P states, whose functions are $O(v^5)$, which are themselves suppressed with respect to 1S states, $O(v^3)$. 

For the fragmentation functions we draw upon the result of Cho and Wise ~\cite{chowise}, who calculated the production cross section of a $1 \an 1D2$ state at the Tevatron. They observed that despite the aforementioned kinematic suppression, large numerical prefactors in the amplitudes implied that such states could be produced in large enough measure to be observed in prompt production. These authors argue that color octet contributions are subleading with respect to color singlet in the heavy quark velocity expansion and thus are neglected. We use their result for the gluon fragmentation function $D_{g\to 1\an 1D2^{(h)}}(z;\mu)$ which describes the production of a $\an 1D2$ quarkonia with helicity $h$, quark longitudinal momentum fraction $z$, and renormalisation scale $\mu$. We compute the production cross section,
\begin{equation}
\frac{d\sigma}{d p_\perp}(p\bar p\to 1\an 1D2+{\rm All})=\sum_{h=0}^{2}\int_0^1 dz 
\frac{d\sigma}{d p_\perp}(p\bar p\to g(p_\perp/z)+{\rm All};\mu)\times D_{g\to 1\an 1D2^{(h)}}(z;\mu)
\end{equation}
using recent gluon distribution functions. We find that the ratio between the MRSD0 gluon distribution functions used in~\cite{chowise} and the most recent MSTW20008NNLO set amounts to about a factor of 0.7 in the most relevant Bjorken $x$ region, $x\simeq \sqrt{\hat s/s}$, which at Tevatron energies is  $x\simeq \sqrt{M_\perp^2/1960^2}\simeq 0.02$ , where $M_\perp=\sqrt{M^2+p_\perp^2}$.
Indeed we have $M=3872$~MeV, $p_\perp\gtrsim 5$~GeV, and $|y|\leq 0.6$ to fulfil the kinematical cuts used in the CDF analysis. Here the factorization scale $\mu$ is set $\mu=M_\perp$.

The integrated prompt cross section we find over the interval $p_\perp\geq 5$~GeV is 
\begin{equation}
\sigma (p\bar p\to 1\an 1D2+all)=0.6~{\rm nb},
\label{charmxsect}
\end{equation}
some 50 and 120 times smaller than the estimated experimental cross section in 
Eq. (\ref{exptvalue}). It is difficult, therefore, to reconcile the observed production cross section of the $\XX$ with the expectations for a $1\an 1D2$ state. As we shall see, the mass of the $X(3872)$ is also inconsistent with the 
$1\an 1D2$ charmonium assignment.

\section{A relativistic hadron string model in the heavy-quark limit}\label{stringmodel}
We turn our attention now to the question of the expected mass of a $1\an 1D2$ state. Typically quark potential models predict a $1\an 1D2$ state lying some 50-100MeV lighter than the $\XX$ mass \cite{Barnes:2003vb}. In the following we will consider the mass of an $1\an 1D2$ state in an evolution of the Chew-Frautschi string model (in the version described by Selem and Wilczek~\cite{selem}) extended to accommodate heavy quarks. The string ends carry spin and electric charge, which allows us to include spin-orbit and magnetic moment interactions.  

As described in~\cite{selem} (see also~\cite{huang}), the energy of a relativistic string with tension $T$
(=$dE/dr$ in the rest frame of a segment $dr$) spinning with angular velocity $\omega$ is
\begin{equation}
{\cal E}=m_1\gamma_1+m_2\gamma_2+\frac{T}{\omega}\int_{0}^{\omega r_1}\frac{dv}{\sqrt{1-v^2}}+\frac{T}{\omega}\int_{0}^{\omega r_2}\frac{dv}{\sqrt{1-v^2}}
\label{stringen}
\end{equation}
with
\begin{equation}
\label{gamma0}
\gamma_i= \frac{1}{\sqrt{1-(\omega r_i)^2}}
\end{equation}
where $r_i$ is the distance of the particle $i$ ($i=1,2$) with mass $m_i$ from the rotation axis.
Analogously we can write the orbital angular momentum of the spinning string as
\begin{equation}
L=m_1\omega r^2_1\gamma_1+m_2\omega r^2_2\gamma_2+\frac{T}{\omega^2}\int_{0}^{\omega r_1}\frac{v^2\,dv}{\sqrt{1-v^2}}+\frac{T}{\omega^2}\int_{0}^{\omega r_2}\frac{v^2\,dv}{\sqrt{1-v^2}}
\label{stringorb}
\end{equation}
which derives from $dL =\omega dI=\omega\; (v^2/\omega^2) \;d{\cal E}$. Now, recall that the four-force is $G^\mu=dp^\mu/ds$, $ds$ being the four-interval. Then $\pmb{G}=1/\sqrt{1-v^2}\;d\pmb{p}/dt=m \omega^2\gamma^2 r\; \pmb{j}$, and $\pmb{j}$ is in the centripetal direction. Thus we have  
\begin{equation}
\label{tension}
T=m_1\gamma_1^2\omega^2 r_1=m_2\gamma_2^2\omega^2 r_2
\end{equation}
Using Eq.~(\ref{gamma0}) and~(\ref{tension}), one can write $\gamma_i$ and $r_i$ as  functions of $T$, $m_i$ and $\omega$
as
\begin{eqnarray}
\label{gamma}
&& \gamma_i=\sqrt{\frac{1}{2} + \frac{1}{2} \sqrt{1 + 4 \left(\frac{T}{m_i\omega}\right)^2}}\\
&& r_i=\frac{1}{\omega}\sqrt{1-\frac{1}{\gamma_i^2}}
\label{erre}
\end{eqnarray}

Plugging (\ref{gamma}) and (\ref{erre}) inside the expressions of ${\cal E}$ and $L$ one can obtain 
the energy and the orbital angular momentum of the system in terms of the $m_1$, $m_2$, $T$, and $\omega$:
\begin{eqnarray}
\label{eandorb2}
&&{\cal E}={\cal E}(m_1,m_2,T,\omega)\\
&&L=L(m_1,m_2,T,\omega)
\end{eqnarray}
In Appendix~\ref{app:limit} we compute the expressions given above 
in the limit of equal and heavy masses $m_1=m_2=M$ with $T/M\omega<<1$, or equivalently, $TR<<M$. In this limit,
\bea
{\cal E}&\approx &2M+\frac{3RT}{2}\label{Elimit}\\
L&\approx&\sqrt{\frac{MTR^3}{2}}\label{Llimit}
\eea
with $R=r_1+r_2$.
In the heavy quark limit, therefore, the orbitally excited mesons have a fixed radius
\be
R\approx\left(\frac{2L^2}{MT}\right)^{1/3}\label{Rlimit}
\ee
and there is a relation connecting the energy and the orbital angular momentum: 
\begin{equation}
{\cal E}(L)\approx 2M+3\left(\frac{T^2L^2}{4M}\right)^{1/3}.
\label{highell}
\end{equation}
The string tension $T$ is related to that of Regge phenomenology $\sigma$ by $T=\sigma/2\pi$. This is in our picture the energy of a heavy quark-antiquark pair with $L\ne 0$ orbital angular momentum. This is a characteristically different Regge trajectory from that of light quarks  $E\sim \sqrt{TL}$, which can also be obtained from the above parametric equations \cite{selem}. It has recently been demonstrated that the spectra of bottomonia does not fit with the standard Regge trajectory \cite{Gershtein:2007nj}, and indeed we shall argue in this paper that the trajectory of Eq. (\ref{highell}) is the correct one to fit the data.

We discussed this method in~\cite{cotugno} and later found that the expression (\ref{highell}) had been previously  mentioned in the literature \cite{Solovev:1999ua}, but not confronted directly with the meson spectrum. In Ref.~\cite{cotugno} we applied it to the spectrum of orbitally excited tetraquark states in a diquark-antidiquark model. We note with interest that the same result, Eq. (\ref{highell}), arises in the Bohr-Sommerfeld quantization of the standard Cornell potential used in quark model calculations \cite{Brau:2000st}. It shall come as no surprise, then, that the results presented here do not differ drastically from those of other models. 

We use Eq. (\ref{highell}) to predict the masses of the orbitally excited mesons with zero quark spin, yielding for the P- and D-wave spin singlets,
\bea
&M(\an 1P1)&=2M+\Delta,\label{1P1}\\
&M(\an 1D2)&=2M+2^{2/3}\Delta,\label{1D2}
\eea
where $\Delta=3\left({T^2}/{4M}\right)^{1/3}$. For the states with nonzero spin, we improve Eq. (\ref{highell}) by considering that at the ends of the string there is a heavy quark-antiquark pair carrying electric and color charge and magnetic moment, which causes mass splittings due to spin-spin, spin-orbit, and tensor interactions.  We thus write the mass formula for states with given total spin $S$, orbital angular momentum $L$, and total angular momentum $J$:
\be
M(^{2S+1}L_J)=2M+\Delta\; L^{2/3}
+M_{SS}\ev{2\pmb{S}_{q}\cdot \pmb{S}_{\overline q}}
+M_{LS}\ev{\pmb{L}\cdot\pmb{S}} 
+M_{\mu\mu}\ev{\pmb{S}^2-3\left(\pmb{S}\cdot\pmb{n}\right)^2}
\label{mass}
\ee
where here $\pmb{S}_q$ and $\pmb{S}_{\overline q}$ are the quark and antiquark spins coupled to $\pmb S$, and the unit vector $\pmb n=\pmb{R}/R$, with $\pmb{R}$ the vector connecting the quark and the antiquark. The mass shifts $M_{SS}$, $M_{LS}$ and $M_{\mu\mu}$ for spin-spin, spin-orbit and tensor interactions are the analogues of the interaction potentials used in quark potential model calculations. The difference is that in this semiclassical approach, instead of taking matrix elements of these $R$-dependent functions between meson wave functions we can evaluate the expectation values directly using the fixed value of the meson radius. We shall derive explicit expressions for these mass shifts later in the paper; first we investigate some general predictions of the model arising from Eq. (\ref{mass}). 

We assume that the spin-spin interaction splits only states with zero orbital angular momentum, for which the $q\bar q$ pair tends to be at much lower average distance. Assuming that $M_{SS}$ is common to both $\an 1S0$ and $\an 3S1$ states, the masses of these states are
\bea
M(\an 1S0)&=&2M-\frac{3}{2}M_{SS}\label{1S0}\\
M(\an 3S1)&=&2M+\frac{1}{2}M_{SS}\label{3S1}
\eea
Equations~(\ref{1P1})-(\ref{1D2})  and~(\ref{1S0})-(\ref{3S1}) yield a single linear relation among the masses
\be
M(\an 1D2)=2^{2/3}M(\an 1P1)+\frac{1-2^{2/3}}{4}\left(M(\an 1S0)+3M(\an 3S1)\right)
\label{linear}
\ee
and a parameter-independent prediction of the mass of the $\an 1D2$ state in terms of the masses of the $h_c$, $J/\psi$ and $\eta_c$,
\be
M(\an 1D2)\simeq 3795~{\rm MeV}.
\label{massprediction}
\ee
The predicted mass is in reasonable agreement with other models, as we discuss later, but is considerably lower than the mass of the $\XX$. In what follows we refine the model, adjusting the parameters to fit also the $\chi_c$ states, but ultimately our predicted $M(\an 1D2)$ changes very little. Thus, our conclusion is that the $\XX$, whose mass exceeds our prediction by some 80 MeV, is difficult to reconcile with a $\an 1D2$ interpretation.

In what follows we verify that the application of the same mass formula in the bottomonia sector is in remarkable agreement with data, supporting the validity of our prediction. It is not possible to apply the formula directly, since the bottomonia $\an 1P1$ and $\an 1D2$ states have not been observed; instead we exploit the relation between the masses of these spin-singlet states and the spin-triplet members $\an 3P{0,1,2}$ and $\an 3D{1,2,3}$ of the same families.

The spin-orbit and tensor terms split the states with spin-one and nonzero $L$, although we will assume, as is usual, that for these states spin-spin contact interactions are negligible. Applying Eq. (\ref{mass}) to the P-wave family yields the following expressions for the masses of the triplet states in terms of the mass of the singlet $\an 1P1$ and the shifts due to spin-orbit and tensor splittings:
\bea
M(\an 3P0)&=&M(\an 1P1)-2M_{LS}^{(\uP)}+2M_{\mu\mu}^{(\uP)}\\
M(\an 3P1)&=&M(\an 1P1)- M_{LS}^{(\uP)}- M_{\mu\mu}^{(\uP)}\\
M(\an 3P2)&=&M(\an 1P1)+ M_{LS}^{(\uP)}+\frac{1}{5}M_{\mu\mu}^{(\uP)}
\eea
where we have labeled the mass shifts $M_{LS}$ and $M_{\mu\mu}$ with a superscript to denote that they are dependent on the meson radius $R$, and hence the partial wave. Eliminating these terms yields a linear relation among the masses of the four members of the P-wave family,
\be
M(\an 1P1)=\frac{1}{9}\left(M(\an 3P0)+3M(\an 3P1)+5M(\an 3P2)\right).\label{pmass}
\ee

The accuracy of this relation can be tested in the charmonia sector, where all four states have been observed. Using the $\chi_{c0}$, $\chi_{c1}$ and $\chi_{c2}$ states as input, one predicts $M(h_c)=3525.30\pm 0.1\textrm{ MeV}$, in remarkable agreement with the experimental mass $M(h_c)=3525.67\pm 0.32 \textrm{ MeV}$ \cite{Amsler:2008zzb}. Reference~\cite{Voloshin:2007dx} uses recent high precision data on the $h_c$ mass and finds that it is in even more striking agreement with this prediction. We note in passing that the good agreement also justifies the neglect of spin-spin interactions for nonzero $L$ which, if present, would modify the relation (\ref{pmass}).

One naturally expects that the equivalent relation ought to hold in the bottomonia sector, and several authors have used it to predict the mass of the as-yet unobserved $h_b$ \cite{Godfrey:2002rp}. With the current masses of the $\chi_{b0}$, $\chi_{b1}$ and $\chi_{b2}$ we get 
\be
M(h_b)=9899.87\pm 0.59 \textrm{ MeV}
\label{hb}
\ee
Recent data \cite{delAmoSanchez:2010kz} on the masses of the $\an 3D{1,2,3}$ states of the upsilon sector allow us to apply the analogous approach, arriving at a new prediction for the mass of the as-yet unidentified $\an 1D2$ state. The mass formulas
\bea
M(\an 3D1)&=& M(\an 1D2)- 3M_{LS}^{(\uD)} + M_{\mu\mu}^{(\uD)}\\
M(\an 3D2)&=& M(\an 1D2)-  M_{LS}^{(\uD)} - M_{\mu\mu}^{(\uD)}\\
M(\an 3D3)&=& M(\an 1D2)+ 2M_{LS}^{(\uD)} +\frac{2}{7} M_{\mu\mu}^{(\uD)}
\eea
lead to the linear relation
\be
M(\an 1D2)=\frac{1}{15}\left(3M(\an 3D1)+5M(\an 3D2)+7M(\an 3D3)\right)
\ee
and we thus predict the mass of the unobserved $\an 1D2$ state, 
\be
M(\eta_{b2})=10165.84\pm 1.8\textrm{ MeV}.
\label{etab2}
\ee

We are now in a position to test the validity of our mass formula (\ref{linear}) in the bottomonia sector. We use as input the experimental values for the $\eta_b$ and $\Upsilon$ masses, and for the $h_b$, encouraged by the remarkable accuracy of the corresponding prediction in the charmonia sector, we use the center of gravity mass in Eq. (\ref{hb}). The mass relation (\ref{linear}) yields a predicted $\an 1D2$ mass of $M(\eta_{b2})=10168.72^{+1.4}_{-1.8}$ MeV, in striking agreement with the above value extracted from the experimental data, and therefore we can be confident in the reliability of the corresponding prediction for the $\an 1D2$ charmonia state.

\begin{figure}
\includegraphics[width=12truecm]{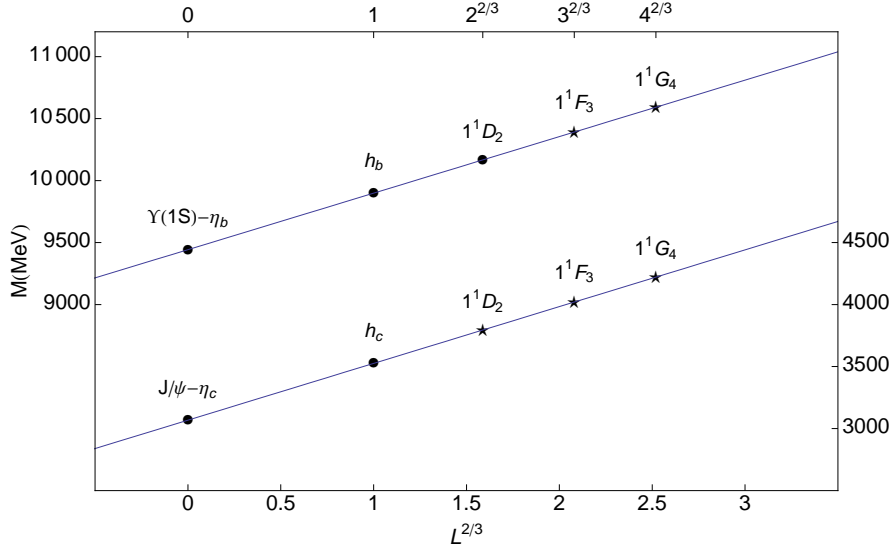}
\caption{Regge trajectories for bottomonia (upper left) and charmonia (lower right) states. The closed circles indicate experimental masses (or those of the center of gravity where appropriate) for a given $L$; stars indicate predictions for unobserved states.}
\label{fig:regge}
\end{figure}

The remarkable accuracy of the model can be seen by plotting the Regge trajectory of equation (\ref{highell}). For the center of gravity of the 1S states ${\cal E}(L=0)=2M$ we use Eqs. (\ref{1S0}) and (\ref{3S1}),
\be
{\cal E}(L=0)=\frac{1}{4}\left(3M(\Upsilon)+M(\eta_b)\right)=9442.45^{+1.20}_{-1.07}\textrm{ MeV}
\ee
and for ${\cal E}(L=1)$ and ${\cal E}(L=2)$ we use the values (\ref{hb}) and (\ref{etab2}) extracted from the data. In the upper part of Fig.~\ref{fig:regge} we plot these three data points in the $\cal {E}$-$L^{2/3}$ plane, and one sees immediately the accuracy with which the model fits the data. On the same plot (indicated by stars) we show our prediction for the masses of the higher $L$ spin-singlet states, $M(\an 1F3)=10391$~MeV and $M(\an 1G4)=10591$~MeV, the identification of which would be a good test of our model. Both of these states should be narrow since they lie below the threshold for $B^*\bar B$, the lightest open flavor pair to which they can decay.

In the lower part of the same figure we plot the corresponding Regge trajectory for the $c\bar c$ system. Here we have only two data points: ${\cal E}(L=0)$, which we determine from the masses of the $\Jp$ and $\eta_c$ as above, and ${\cal E}(L=1)=M(h_c)$. We indicate by stars our predictions for $M(\an 1D2)=3795$~MeV, $M(\an 1F3)=4020$~MeV and $M(\an 1G4)=4221$~MeV, although the masses of the latter pair may not be as reliable as in the bottomonia case because of the coupling to the open $D^*\bar D$ threshold.

A striking feature of Fig.~\ref{fig:regge} is that the slopes of the trajectories for $b\bar b$ and $c\bar c$ are almost identical. Since the slope is given by $3(T^2/4M)^{1/3}$, this implies that the effective string tension scales with the square root of the quark mass, in contrast to many potential models in which a common string tension is used for different quark masses; we determine the best fit values $T_b=0.257$~GeV$^{2}$ and $T_c=0.148$~GeV$^{2}$. From the intercept of the trajectories we determine the quark masses $M_b=4721$~MeV  and $M_c=1533$~MeV.

We return now to the derivation of the mass shifts $M_{SS}$, $M_{LS}$, and $M_{\mu\mu}$ in terms of the parameters of the string model. In a potential model approach the interaction potential for the spin-spin term is
\be
V_{SS}(\pmb R)=\kappa_{c\bar c}\delta^3(\pmb R).
\ee
\noindent The corresponding mass shift is thus proportional to the square of the wave function at the origin, assuming that the magnetic coupling and the wave function at the origin do not depend on the total spin of the pair
\begin{equation}
\label{ss}
{M}_{SS}=\ev{V_{SS}(\pmb R)}=\kappa_{c\bar{c}}\;\left|\psi_{_{1{\rm S}}}(0)\right|^2.
\end{equation}
\noindent  In our model we treat $\kappa'=\kappa|\psi(0)|^2$ as a free parameter. The mass shift $M_{LS}$ is due to the coupling of the magnetic moment of one quark  to the magnetic field created by the moving charge of the other quark; this is the essence of the spin-orbit coupling, 
\begin{equation}
M_{LS}=A\frac{1}{R}\frac{\partial {\cal E}(R)}{\partial R}
\label{modelso}
\end{equation}
where $A$ is a constant to be determined and ${\cal E}$ is the energy of the string as in Eq.~(\ref{stringen}) with $R=r_1+r_2$, an approximately linear function as shown in Eq.~(\ref{Elimit}). The tensor mass shift $M_{\mu\mu}$ arises from
the interaction between the magnetic moment of one of the two quarks
with the static magnetic field generated by the other one, 
\be
\label{tenso}
M_{\mu\mu}=B\frac{1}{R^3}
\ee
where the constant $B$ is again to be determined.

Following standard electromagnetism, one can parametrize $A$ and $B$ in terms of the gyromagnetic factor
 and of the charge of the heavy quarks bound to form the meson. The interaction between the magnetic dipole and the magnetic field generated from a moving charge
carries a factor $ge/2m^2$, 
whereas the Thomas precession gives $-e/2m^2$. 
As for the tensor coupling $B$ we expect it to be related to the product of the Bohr magnetons of the quark and the antiquark. Thus, one is led to write
\be
\label{evvai}
A=\frac{(g_{_Q}-1)e_{_Q}}{2m_{_Q}^2},\qquad B=-\left(\frac{g_{_Q} e_{_Q}}{2m_{_Q}}\right)^2.
\ee
As we are dealing with a constituent quark model we do not expect the $g_Q$ to be comparable to those of a pointlike structureless particle. Thus, we leave $g_Q$ as a free parameter
and, as expected, we find best fit values larger than 2. We use $e_c=2e/3$ and $e_b=e/3 $.

\begin{figure}
\begin{center}
\begin{minipage}[t]{7truecm} 
\centering
\includegraphics[width=7truecm]{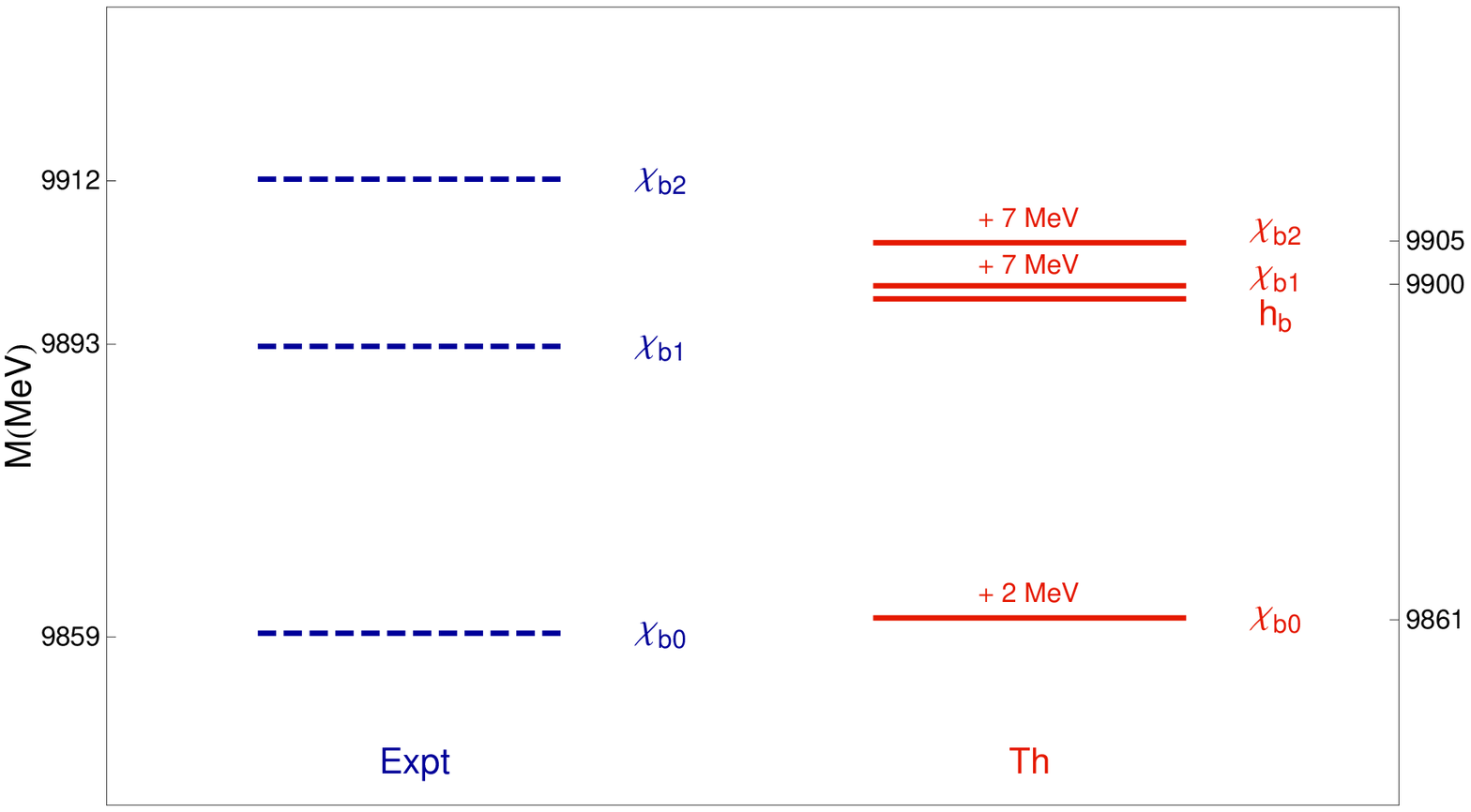}
\caption{Results of the fit to the $L=1$ bottomonia with the parameters in Eq.~(\ref{resb}).}
\label{fig:fitchibj}
\end{minipage}
\hspace{1truecm} 
\begin{minipage}[t]{7truecm}
\centering
\includegraphics[width=7truecm]{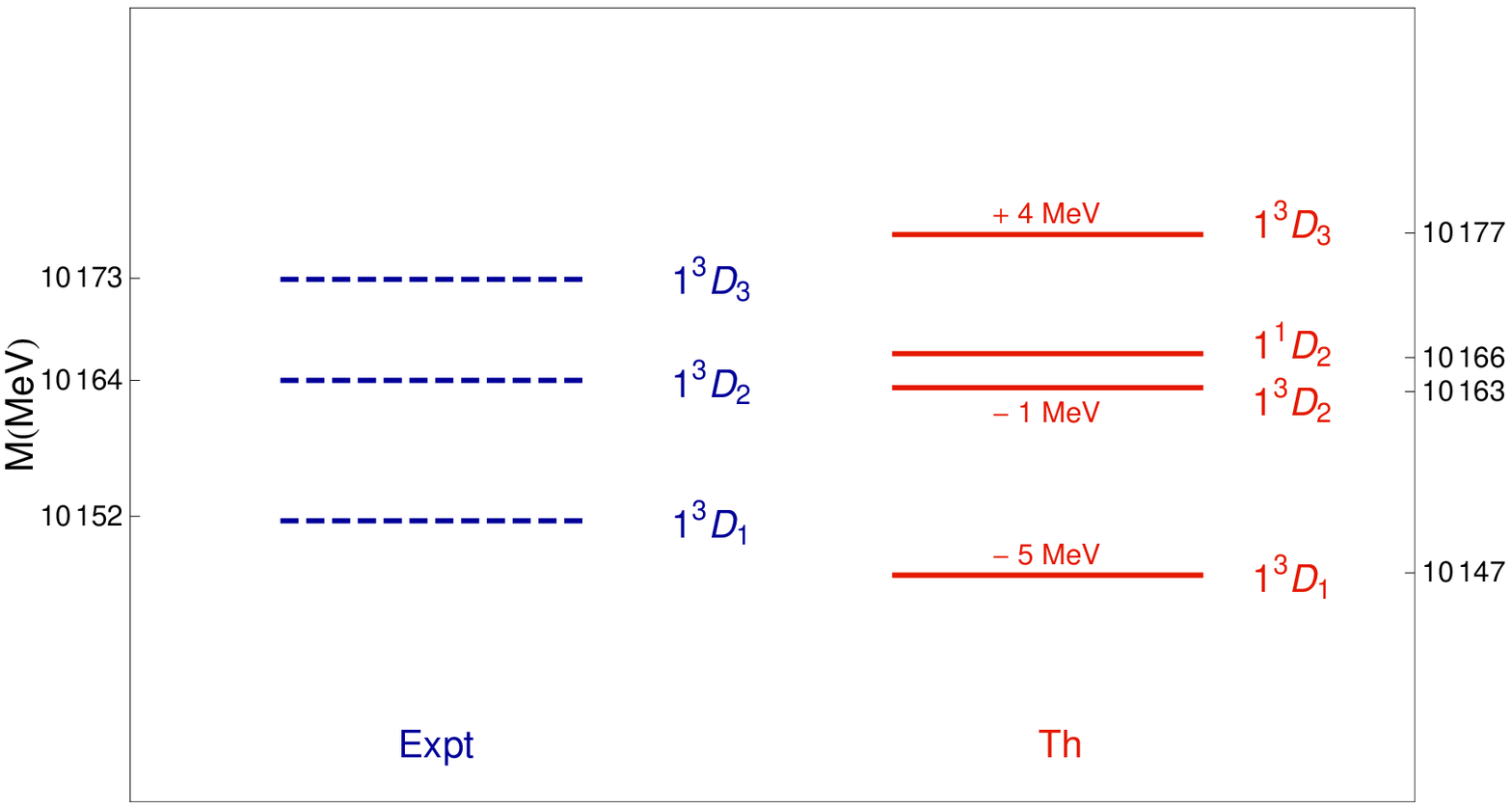}
\caption{Results of the fit to the $L=2$ bottomonium states with the parameters in Eq.~(\ref{resb}).}
\label{fig:fitupsilon1D}
\end{minipage}
\end{center}
\end{figure}

We test the validity of our model on the charmonia and bottomonia spectra with $L=1,2$. In the heavy quark limit we expect the model to work better at fitting the bottomonia states, and we fit the $\eta_b$, $\Upsilon$, $\chi_{b0},\chi_{b1},\chi_{b2}$~\cite{Amsler:2008zzb}, and $\Upsilon(1D_1),\Upsilon(1D_2),\Upsilon(1D_3)$~\cite{delAmoSanchez:2010kz}.  The splitting between $\eta_b$ and $\Upsilon$ is not a feature of the string model $per$ $se$, since it is controlled by the parameter $\kappa_b'$ which is not correlated with any other masses. The masses of the remaining states are controlled by two parameters ($T_b$ and $g_b$), and we find the best fit values
\be
\label{resb}
\begin{cases}
\,T_b=0.258~{\rm GeV}^{2}\\
\,g_b = 11.5\\
\end{cases}
\ee
The string tension is close to the value used to fit Regge trajectories of light mesons ($T\approx 0.175~{\rm GeV}^{2}$) and differs only very slightly from the value obtained earlier by fitting the spin-averaged masses. The resulting spectrum is summarized in Table~\ref{tab:summary1B}, where we also compare our results with the predictions obtained in the potential models of Refs.~\cite{Eichten:1980mw} and~\cite{Godfrey:1985xj}. The agreement is remarkable. The splitting between the center of gravities of the S-, P-, and D-wave sectors has already been confronted with data; the new feature is that with the same parameters we are also able to describe the splitting among the $\an 3P{0,1,2}$ and $\an 3D{1,2,3}$ states. A comparison of the theoretical and experimental spectra is presented in Figs.~\ref{fig:fitchibj} and~\ref{fig:fitupsilon1D}. 
\begin{table}[!h]
  \begin{tabular}{|| lc|c|c|c|c||}
    \hline
    & $M_{exp}~({\rm{MeV}})$ & $\sigma(M_{exp})~({\rm{MeV}})$ &$M_{th}~({\rm MeV})$ &  $M_{th}~({\rm MeV})$ \textbf{\cite{Eichten:1980mw}} & $M_{th}~({\rm MeV})$ \textbf{\cite{Godfrey:1985xj}} \\
    \hline
    $\eta_b$      & $9391$  &$3.2$ & $\mathbf {9388}$  & $9366$ & $9400$\\
    \hline
    $\Upsilon(1S)$     & $9460$  & $0.26$& $\mathbf{9460}$ &  $9460$ & $9460$\\
     \hline
    $h_{b}$     &   &    &$\mathbf{9898}$  & $9924$ & $9880$\\
    \hline  
    $\chi_{b0}$     & $9859$ &$0.5$ & $\mathbf{9861}$  & $9888$ & $9850$\\
    \hline
    $\chi_{b1}$     & $9893$ &$0.4$ & $\mathbf{9900}$  & $9913$ & $9880$\\
    \hline
    $\chi_{b2}$      & $9912$  &$0.4$& $\mathbf{9905}$ & $9939$ & $9900$\\
        \hline
    $\Upsilon(1\an 1D2)$     &    &  & $\mathbf{10166}$ & $10166$ & $10150$\\
    \hline
        $\Upsilon(1^3D_1)$    & $10152$ &$1$  & $\mathbf{10147}$ & $10153$ & $10140$\\
        \hline
    $\Upsilon(1^3D_2)$     & $10164$ &$0.9$   & $\mathbf{10163}$  & $10163$ & $10150$\\
        \hline
    $\Upsilon(1^3D_3)$     & $10173$  &$1$ & $\mathbf{10177}$  & $10174$ & $10160$\\
\hline
  \end{tabular}
\caption{Theoretical values of the masses obtained by the fit  with the parameters in Eq.~(\ref{resb}) compared to the experimental values and to other theoretical determinations~\cite{Eichten:1980mw,Godfrey:1985xj}.}
\label{tab:summary1B}
\end{table}

The same work can be done for charmonia. In this case a priori we do not expect such a remarkable agreement with data.
Still, as we will see, the agreement with the $L=1$ states is rather good.
We include in our fit the following charmonia~\cite{Amsler:2008zzb}: $\eta_c$, $\Jp$, $h_{c}$, $\chi_{c0},\chi_{c1},\chi_{c2}$, and $\psi(3770)$, which we identify with the $1^{3}D_{1}$ state.
The best fit is obtained with the following values for the parameters:
\be
\label{res}
\begin{cases}
\,T_c=0.147~{\rm GeV}^{2}\\
\,g_c = 5.7\\
\end{cases}
\ee
Here the agreement of the value obtained for the string tension with the one used for light mesons is better than in the $b\bar{b}$ case, and again it differs very little from the value obtained in the spin-averaged case.
It is worth saying that performing the same fit procedure numerically with the full expression for the energy, {\it i.e.}, without any expansion, and orbital momentum [see Eqs.~(\ref{stringen}) and (\ref{stringorb})], the  best fit parameters are essentially the same as the ones quoted above, both for the charmonium and the bottomonium sectors.

\begin{figure}
\begin{minipage}[t]{7truecm} 
\centering
\includegraphics[width=7truecm]{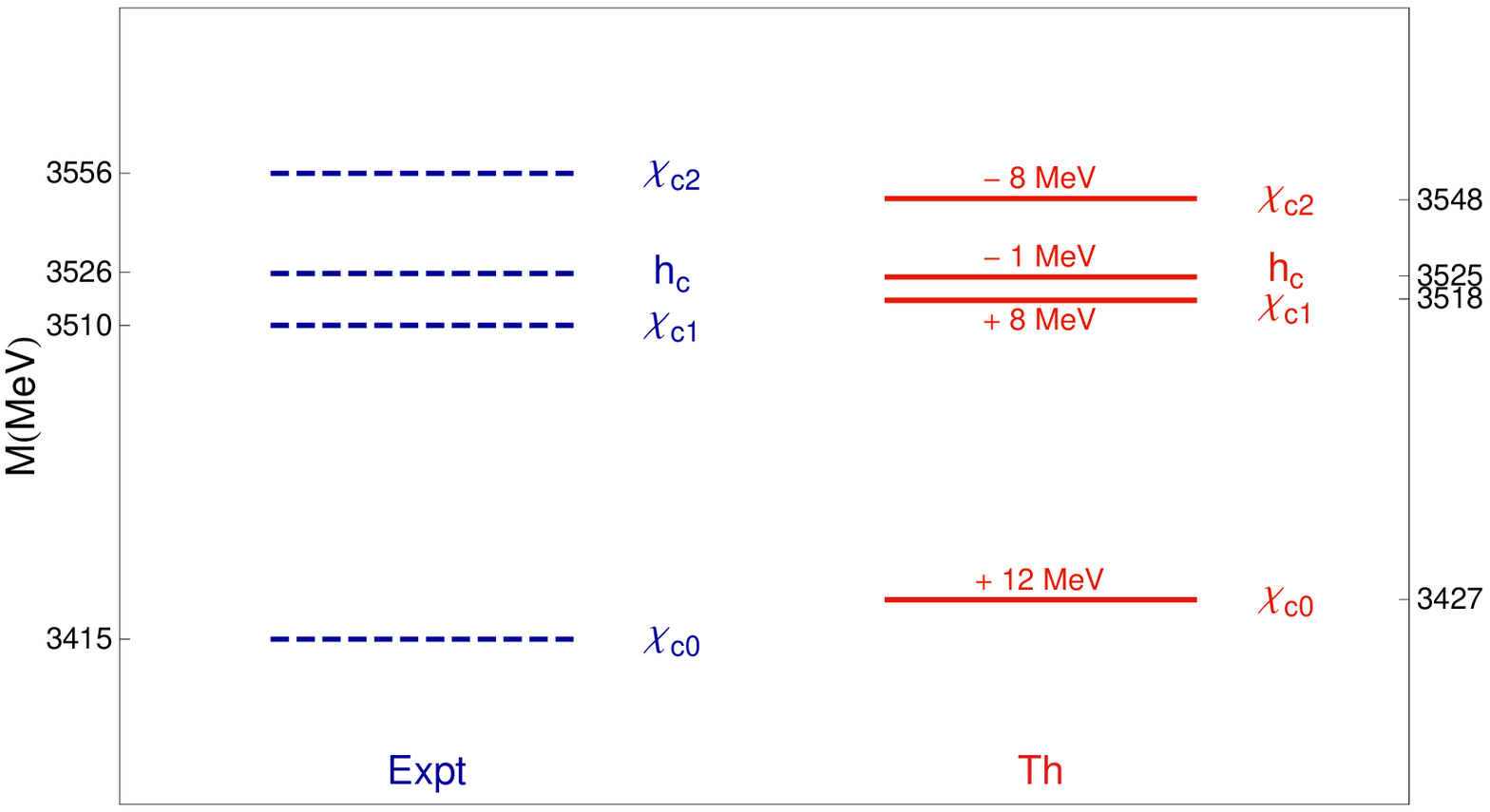}
\caption{Results of the fit to the $L=1$ charmonium states with the parameters in Eq.~(\ref{res}). The agreement between data and the results of our string model are remarkable in the $L=1$ sector.}
\label{fig:fitchicj}
\end{minipage}
\hspace{1truecm} 
\begin{minipage}[t]{7truecm}
\centering
\includegraphics[width=7truecm]{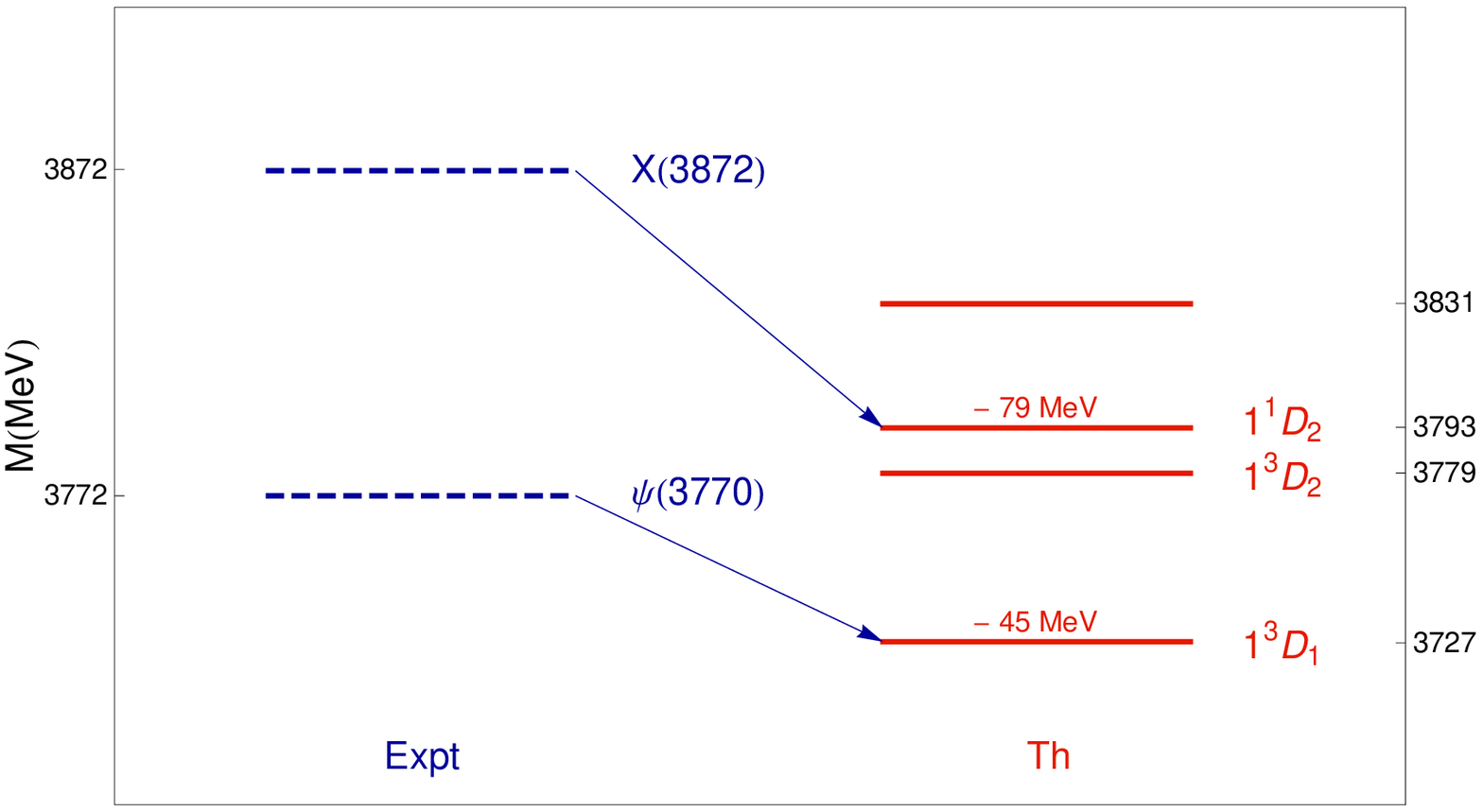}
\caption{ The same as in Fig.~\ref{fig:fitchicj} but for the $L=2$ states. We tentatively identify the $X(3872)$ with the $1\an 1D2$ charmonium state and the $\psi(3770)$ with the $1\an 3D1$.}
\label{fig:fitelle2}
\end{minipage}
\end{figure}

The experimental masses \cite{Amsler:2008zzb} are compared to the results of the fit in Table~\ref{tab:summary1}, where we again compare with the predictions of Refs.~\cite{Eichten:1980mw,Godfrey:1985xj}, and in Figs.~\ref{fig:fitchicj} and~\ref{fig:fitelle2}. The agreement is very good, although unlike Refs.~\cite{Eichten:1980mw,Godfrey:1985xj} our model fails to reproduce the mass of the $\psi(3770)$, which is around $45~{\rm MeV}$ above our predicted value. The effect of the mixing with the nearby $\psi'(3686)$ could be advocated as a possible 
resolution of this mismatch. However, since the mixing angle between the two states is only $\phi=(12\pm 2)^{\circ}$~\cite{Rosner:2001nm,Rosner:2004wy}, 
it could account for an upward shift of at most $\sim 5~{\rm MeV}$ with respect to the fitted value, giving the $\psi(3770)$ at $3735~{\rm MeV}$.
\begin{table}[!h]
  \begin{tabular}{|| lc|c|c|c|c|c|c||}
    \hline
    & $M_{exp}~({\rm{MeV}})$ & $\sigma(M_{exp})~({\rm{MeV}})$ &  $M_{th}~({\rm MeV})$  &  $M_{th}~({\rm MeV})$ \textbf{\cite{Eichten:1980mw}} & $M_{th}~({\rm MeV})$ \textbf{\cite{Godfrey:1985xj}} \\
    \hline
    $\eta_c$      & $2980$   &$1$ & $\mathbf {2981}$  & $2965$ & $2970$\\
    \hline
    $J/\psi$     & $3096$  & $0.1$ & $\mathbf{3097}$ &  $3095$ & $3100$\\
     \hline
    $h_{c}$     & $3526$     & $0.32$ & $\mathbf{3525}$  & $3525$ & $3520$\\
    \hline  
    $\chi_{c0}$     & $3415$  & $0.31$   & $\mathbf{3427}$  & $3415$ & $3440$\\
    \hline
    $\chi_{c1}$     & $3510$    & $0.07$ & $\mathbf{3518}$  & $3508$ & $3510$\\
    \hline
    $\chi_{c2}$      & $3556$  & $0.09$  & $\mathbf{3548}$ & $3555$ & $3550$\\
        \hline
    $X(3872)[1\an 1D2]$     & $3872$ & $0.24$  & $\mathbf{3793}$ & $3810$ & $3840$\\
    \hline
    $\psi(3770)\:[1^3 D_1]$     & $3772$ & $0.35$  & $\mathbf{3727}$ & $3762$ & $3820$\\
        \hline
    $1^3 D_2$     &    &  & $\mathbf{3779}$  & $3797$ & $3840$\\
        \hline
    $1^3 D_3$     &   &  & $\mathbf{3831}$  & $3840$ & $3850$\\
\hline
  \end{tabular}
\caption{Theoretical values of the masses obtained by the fit  with the parameters in Eq.~(\ref{res}) compared to the experimental values and to other theoretical determinations~\cite{Eichten:1980mw,Godfrey:1985xj}.}
\label{tab:summary1}
\end{table}

Our predicted value of 3793~MeV for the mass of the $1\an 1D2$ state is only slightly different from that which we obtained using the formula (\ref{massprediction}), and our conclusion remains the same: the mass of the $\XX$ is difficult to reconcile with the charmonium interpretation. Indeed it is evident in Figs.~\ref{fig:fitchicj} and~\ref{fig:fitelle2} the most difficult identification to make is that of the $X(3872)$ as a $1\an 1D2$ state. As can be seen in Table~\ref{tab:summary1}, the same is also true of the potential models~\cite{Eichten:1980mw} and~\cite{Godfrey:1985xj}. Reference \cite{Barnes:2003vb} compiles the mass predictions for a variety of different potential models, and five out of the six cases (including the two which we have quoted in this paper) predict the mass of the $1\an 1D2$ is some 50-100~MeV lighter than that of the $\XX$. The exception is that of Fulcher \cite{Fulcher91perturbative} who predicted a $1\an 1D2$ with a mass of none other than 3872~MeV; notwithstanding the remarkable agreement of that model with the $\XX$ mass, we find rather more compelling the broad agreement of our model with the predictions of the majority of other approaches. We note in passing, however, that lattice QCD predicts a somewhat higher $2^{-+}$ mass of $3907\pm 32$~MeV \cite{Dudek07charmonium}, consistent with the $\XX$.

\section{Decays of $\an 1D2$ charmonium}\label{charmoniumdecays}
We have discussed two difficulties with the charmonium interpretation of the $\XX$: its prompt production is too small, while its mass is too large. Another difficulty is the isospin violation in its decays. Indeed the $J/\psi\rho$ and $J/\psi\omega$ modes are observed with the same strength: $\mathcal{B}(X\to J/\psi\omega)/\mathcal{B}(X\to J/\psi\pi^+\pi^-)=1.0\pm0.4~{\rm (stat.)}\pm0.3~{\rm (syst.)}$~\cite{AbeEtAl05evidence}, in obvious contradiction to expectations for a standard $c\bar c$ state which is a pure isoscalar. However the isospin violation is not so severe if one considers the different phase space volumes available to $J/\psi\pi^+\pi^-$ and $J/\psi\pi^+\pi^-\pi^0$ final states. Because of the different decay widths of $\rho$ and $\omega$ one finds the ratio between the $I=1$ and $I=0$ amplitudes $\mathcal{A}_{I=1}/\mathcal{A}_{I=0}\approx 0.5$, consistent with the experimental value. A further possibility is that the $J/\psi\pi^+\pi^-$ and $J/\psi\pi^+\pi^-\pi^0$ modes are fed by rescattering from intermediate $D^*\bar D$ states, and isospin is broken by the mass difference between the neutral and charged states \cite{Suzuki05x(3872)}. 

Historically, the literature on the $1\an 1D2$ state has assumed that it lies below the $D^*\bar D$ threshold in which case it would decay dominantly by radiative transitions, hadronic transitions with pion emission, and annihilation into gluons. Among the radiative transitions, the dominant mode is expected 
to be~\cite{chowise}
\be
X(3872) \to h_c\gamma\to J/\psi \pi^0\gamma, 
\label{accac}
\ee
with a branching ratio of 0.004. 
Using the $\XX$ mass, Ref. \cite{Barnes:2003vb} predicts a $h_c\gamma$ width of 0.460~MeV, somewhat larger than an earlier prediction of 0.278~MeV which assumed a lower mass \cite{EichtenLaneEtAl02b-meson}. Reference \cite{Barnes:2003vb} also predicts a considerable $\psi(3770)\gamma$ mode with a partial width of 0.045~MeV. By contrast, the observed $\Jp\gamma$ and $\psi'\gamma$ are rather at odds with expectations. Because of the orthogonality of the spin and spatial wave functions of the heavy quark pair, these transitions are expected to be small, although the $\psi'\gamma$ would be enhanced due to its 1D component, as it is presumably the orthogonal partner of the dominantly 1D $\psi(3770)$. 
In a recent paper~\cite{JiaSangEtAl10is} the $2^{-+}$
hypothesis has been tested on the radiative decays of $X(3872)$.
Using potential non-relativistic QCD 
(coupled to electromagnetism to describe single photon transitions), they found upper bounds on $\mathcal{B}(1^{1}D_2\to J/\psi(\psi^{'})\gamma)$ which are 1 order of magnitude smaller than the experimental lower bounds obtained from BaBar measurements. 
The estimate for the $1^1D_2\to \psi^{'}\gamma$ takes into account the $S-D$ wave mixing in the $\psi^{'}$ wave function, which turns out to be not sufficient to  accommodate theoretical predictions 
with experimental data.
The approach used in~\cite{JiaSangEtAl10is}, first derived in~\cite{Brambilla:2005zw}, relies on the use of charmonium potential models in order to compute overlap integrals of radial wave functions. The stability of the results is checked using different potentials. 
Furthermore in recent news Belle does not confirm the $X(3872) \to \psi^\prime \gamma$ 
decay~\cite{bellenopsip}, in contrast to the BaBar result with an estimated branching ratio ${\cal B}(X\to \psi^{'}\gamma)\simeq 6\%$~\cite{AubertEtAl09evidence}.

As for the hadronic transitions 
Refs.~\cite{EichtenLaneEtAl02b-meson, Barnes:2003vb} indicate that the dominant hadronic decay mode with two pions in the final state is $1\an 1D2\to \eta_c\pi\pi$, and the latter authors predict a partial width of $0.21\pm 0.11$~MeV. 
As described in~\cite{Gottfried:1977gp}, this result is obtained using a multipole expansion of the color gauge field. In atomic physics the analogous process is the emission of atomic electrons or $e^+e^-$ pairs by nuclei.
This expansion leads to some selection rules which at leading order imply that 
quark spin is conserved. Thus the operator responsible for the transition connects quarkonium states with the same spin $S$, namely $1\an 1D2$ and $1\an 1S0$, the $\eta_c$.
The observation or the absence of this decay would be crucial and thus we encourage the experimental search in this direction.

Reference \cite{EichtenLaneEtAl04charmonium} considered the $\ddsn$ decays of a heavier $1\an 1D2$ and predict a partial width of 0.03~MeV for a state with mass 3872~MeV, increasing to 1.7~MeV if it were somewhat higher at 3880~MeV; experimentally, the $\XX$ is known to lie much closer to 3872~MeV and these partial widths are consistent with the experimental width.

\section{The Tetraquark interpretation}\label{4qsection}
In this section we consider the tetraquark option which, unlike the molecule, may still be a viable interpretation of the $\XX$. One advantage of the tetraquark interpretation is that it admits a natural solution to the isospin violation issue, due to suitable mixing between the diquark flavor states \cite{4q}.
To form a diquark-antidiquark state with $2^{-+}$ quantum numbers we need a unit of orbital angular momentum between the two diquarks (the diquarks have positive parity), and at least one out of the two diquarks must have spin 1. 
The existence of such a state therefore implies the existence of a rich spectrum of partner states, formed of combinations of spin 1 (axial vector) and spin 0 (scalar) diquarks coupled to the $L=1$ orbital angular momentum to give various total angular momenta $J$; in addition, one expects a series of lighter $L=0$ states with positive parity. The spin wave functions and corresponding $J^{PC}$ of $L=0$ and $L=1$ tetraquarks is discussed in Ref.~ \cite{Drenska:2009cd}; we extend their results to states with $J=2$ and 3 and  summarize these in Table~\ref{tab:tetra}. We shall assume, as is usual, that only color (anti)triplet diquarks form stable configurations \cite{Jaffe05exotica}.

\begin{table}%[ht!]
\begin{tabular}{l|cc}
\hline
\hline
$|(S_{cq}\times S_{\bar c \bar q})_{S}\rangle$	
								  &  \hspace{0.5truecm}$L=0$	\hspace{0.5truecm}	&$L=1$	\\
\hline

$\k|(0\times 0)_0>$						& $0^{++}$	&$1^{--}$ \\
$\k|(1\times 1)_0>$						& $0^{++}$	&$1^{--}$ \\
$\sqrt{1/2}\left(\k|(1\times 0)_1>+\k|(0\times 1)_1>\right)$	& $1^{++}$	&$0^{--}1^{--}2^{--}$\\
$\sqrt{1/2}\left(\k|(1\times 0)_1>-\k|(0\times 1)_1>\right)$	& $1^{+-}$	&$0^{-+}1^{-+}2^{-+}$\\
$\k|(1\times 1)_1>$						& $1^{+-}$	&$0^{-+}1^{-+}2^{-+}$\\
$\k|(1\times 1)_2>$						& $2^{++}$	&$1^{--}2^{--}3^{--}$\\
\hline
\hline
\end{tabular}
\caption{The spectrum of $L=0$ and $L=1$ tetraquark states classified according to their spin wave functions (with $S_{cs}$, $S_{\bar c \bar s}$, and $S$ the diquark, antidiquark, and total quark spin, respectively) and $J^{PC}$ quantum numbers.}
\label{tab:tetra}
\end{table}
In general a tetraquark with a given $J^{PC}$ can be formed from one of several different spin wave functions, and we thus expect that the physical states are admixtures of the various possible states of $(S_{cq}\times S_{\bar c \bar q})_S$. For a $2^{-+}$ tetraquark there are two possibilities: an orbital excitation of a vector diquark pair coupled antisymmetrically in spin or a scalar and vector diquark likewise coupled antisymmetrically. One needs also to consider the various flavor combinations; there should exist, for each angular momentum configuration, two neutral and two charged states formed out of $[cu]$ and $[cd]$ building blocks and the corresponding antidiquarks (and a further four strange states if one considers strange diquarks). This presents something of a problem from a phenomenological point of view; if the $\XX$ is a $2^{-+}$ tetraquark state then, $a$ $priori$ at least, it should be accompanied by at least seven partner states, three neutral and four charged. 

It is possible that the data reflects the existence of two neutral states, the lighter state decaying to $\Jp\rho$ and $\Jp\omega$ and the heavier to $\ddsn$, although the experimental interpretation is not clear \cite{AubertEtAl08study,AdachiEtAl08study}. In Ref.~ \cite{MaianiPolosaEtAl07indications} this possibility was discussed in the context of the $\XX$ as a $1^{++}$ tetraquark, where a total of two neutral states and two charged states are expected. Even if there are two states in the data, in the $2^{-+}$ interpretation there would still be two missing neutral states. Moreover, we expect four charged states, for which there is apparently no experimental evidence \cite{Aubert:2004zr}. The problem may not be so severe if one takes account of the mass difference between scalar and vector diquarks, which in the relativistic approach of Ref. \cite{EbertFaustovEtAl08excited} is around 60MeV. In this case the physical $\XX$ could be identified with the state dominated by the scalar-vector combination in Table \ref{tab:tetra}, and the vector-vector combination would be somewhat heavier. If the P-wave coupling to $\Jp\omega$, $\Jp\rho$, and $\ddsn$ blows up rapidly with momentum, then this heavier state may be too broad to be observed; in this case one still needs two neutral and two charged states, instead of four and four.  

We also expect the $\XX$ to be accompanied by a staggering array of partners with different $J^{PC}$: one each of $0^{--}$ and $3^{--}$, two each of $2^{--}$, $0^{-+}$, $1^{-+}$ and $2^{-+}$, and four $1^{--}$. For each $J^{PC}$ we again expect two neutral and two charged members, and possibly also strange states. If the S-wave states also exist one expects, also multiplied fourfold in flavor, two each of $0^{++}$ and $1^{+-}$, and one each of $1^{++}$ and $2^{++}$. In the absence of experimental evidence for these partner states a tetraquark explanation for the $\XX$ is only tenable if one can explain why it, among all the possible configurations, is unique. We shall propose one possibility which is that most of the states are so broad as to be effectively unobservable, while the $\XX$, due the scarcity of allowed hadronic decays and limited phase space, may be among the few states which are narrow and thus observable.

{\bf \emph{Spectrum}}. 
Before addressing such questions we consider the masses we expect for these tetraquark states. For the $L=1$ states there will be presumably be some spin-orbit splittings, although we do not calculate these here. We shall assume, instead, that the states lie around the mass of the $\XX$, somewhere in the $3800-3900$~MeV mass region. Moreover, we assume that the scalar and vector diquarks have equal masses. We fit the diquark mass with the relativistic string model~(\ref{highell}) in such a way that at $3872$~MeV we find the $L=1$ state. This implies
\be
\label{diqm}
m_{[cq]}=1716~{\rm MeV},
\ee
\noindent where we use for the string tension $T$ the value obtained from the fit to the $c\bar{c}$ spectrum. In Ref.~\cite{4q} we discussed the interpretation of the $\XX$ as an S-wave $1^{++}$ state, and correspondingly the diquark mass $m_{[cq]}=1933$~MeV was larger than what we obtain here. 

\begin{figure}
\begin{center}
\includegraphics[width=13truecm]{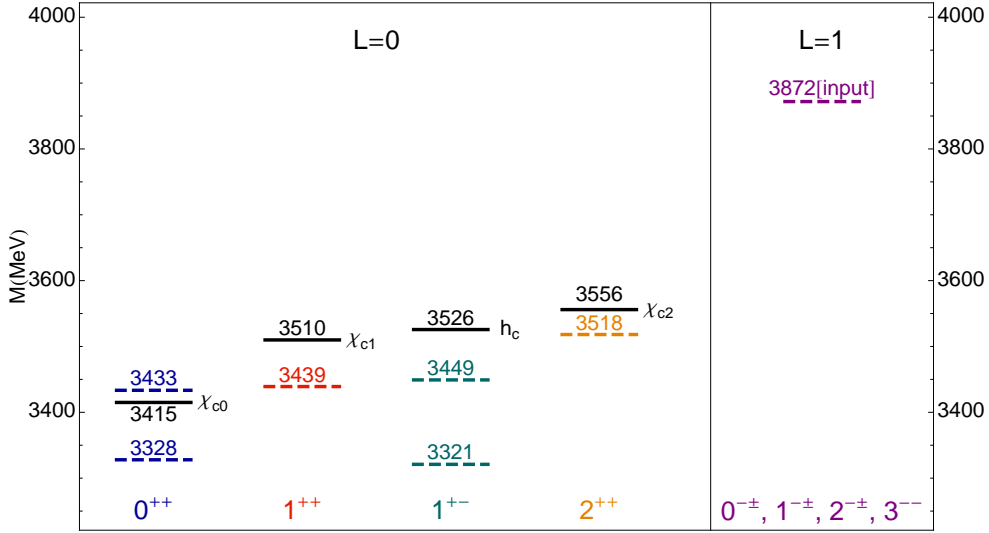}
\end{center}
\caption{Dashed lines represents the full spectrum of the tetraquark $[cq][\bar c\bar q]$ states with $L=0$ and $L=1$, whereas the solid lines are the standard charmonium levels with $L=1$.}
\label{fig:4quark}
\end{figure}

The entire spectrum of $L=0$ states discussed in~\cite{4q} is shifted downwards in mass once the new input value is the $2^{-+}$ $X(3872)$: the average mass will be around $2m_{[cq]}\simeq 3430~{\rm {MeV}}$. Thus a striking prediction of the tetraquark model is that there is a set of positive parity states in the same mass region as the $\chi_{c0,1,2}$ and $h_c$. Their masses can be expressed in terms of the diquark mass and of the magnetic couplings among the four quarks which constitute the state, since at $L=0$ one can assume that only spin-spin interactions are relevant \cite{4q}. The full spectrum of tetraquark states is reported in Fig.~ \ref{fig:4quark}. Among the $L=0$ states the $0^{++}$ tetraquark is very close in mass to the standard charmonium state $\chi_{c0}$, while the others are some $40~{\rm MeV}$  lighter than the charmonia with the same $J^{PC}$.

As for the $L=1$ part of the spectrum, a closer look is required for the $1^{--}$ states. In a previous paper~\cite{cotugno} we proposed a tetraquark interpretation for the vector resonances $Y(4350)$ and $Y_B(4660)$. Motivated by a reanalysis of the Belle data on the $Y_B(4660)$ which shows a preference for the baryon-antibaryon decay mode (the most striking tetraquark signature), these states were identified with $L=1$ tetraquark states with zero and one unit of radial excitation, respectively. 
The masses of the states were obtained using Eq.~(\ref{highell}) with $M=1933~{\rm MeV}$, the value obtained for the diquark mass in~\cite{4q}
in the hypothesis that the $X(3872)$ were a $1^{++}$ state. Reference \cite{EbertFaustovEtAl08excited} considered a similar scheme in a relativistic diquark-antidiquark model and likewise obtained a spectrum of $L=1$ states in the 4200-4300~MeV mass region.

The new value of the diquark mass~(\ref{diqm}) pushes the masses of the $L=1$ states down to the $3800-3900~{\rm MeV}$ region. In this scheme
the $Y(4350)$ and the $Y_B(4660)$ could thus be the two successive radial excitations of the $1^{--}$ state at about $3870$~MeV~\footnote{Observe that using the tension obtained from the fit to the charmonium states, the mass of $L=1$ and $L=3$ tetraquarks would be $3872~{\rm MeV}$ and $4353~{\rm MeV}$ respectively. We choose not to consider the $L=3$ assignment.}.
For a rough estimate of the mass splitting between different radial excitations, we borrow the results obtained for the P-wave charmonia~\cite{Godfrey:2008nc}, which indicate a mass gap from the ground to first radial excitation of around 440~MeV, and from the first to second radial excitation of around 370~MeV. Using this approach, as in~\cite{cotugno}, we predict 
\be
\begin{split}
&M(n_r=1,L=1)=4320~{\rm MeV}\\
&M(n_r=2,L=1)=4690~{\rm MeV}
\end{split}
\ee
The advantage of this assignment is that it explains the suppression of the $Y(4350)\to J/\psi\pi^+\pi^-$ and $Y_B(4660)\to J/\psi\pi^+\pi^-$ decays with respect to the corresponding $\psi^{'}$ decay modes. The $\Jp$ modes should be enhanced by phase space, but may be suppressed due to different radial quantum numbers between the charm quarks in the initial and final states. In~\cite{cotugno} it is shown that the measured spectra indicate that the $\pi\pi$ pair comes from phase space (or, equivalently a $\sigma$) apart from the case of the 
$Y_B(4660)$ where there seems to be a 30\% component due to $f_0(980)$.
As it was done in~\cite{cotugno}, we attempt an explanation of the observed $Y$'s decay pattern describing the S-wave transition $\langle \psi (1S,2S) \mathbbmss{a} |Y\rangle$ as
\begin{equation}
\langle \psi(\eta,q) \mathbbmss{a}(k) |Y(\epsilon,p)\rangle =g \; \epsilon\cdot \eta
\label{matr0}
\end{equation}
where $ \mathbbmss{a}=\sigma,f_0$. 
Interpreting the $Y$'s as  charmoniumlike bound states made up by a diquark and an antidiquark, $g$ can be estimated from the radial overlap integral, 
\begin{equation}
 g\propto \int d^3 r    \;R^{\dagger}_{{(\psi)}}(r)
R_{_{(Y)}}(r)
\label{matr}
\end{equation}

\noindent where $R_{_{(Y)}}(r)$ is the radial wave function for the P-wave $[cq][\bar c \bar q]$ states and $R^{\dagger}_{{(\psi)}}(r)$
refers to the standard $c\bar{c}$ states.
Since our string model provides us with an expression of the interaction energy as a function of the distance between the 
quark and the antiquark in the case of a charmonium state, and the diquark and the antidiquark in the case of a tetraquark,
we solve numerically the radial part of the Schr\"odinger equation using for the potential equation~(\ref{dist}) with the parameters $T$ and $M$ obtained from the charmonia fits. 
We are interested in estimating the following ratio of decay widths:
\begin{equation}
\Gamma_\mathbbmss{a} (Y) \equiv \Gamma(Y\to  \psi(1S) (\pi\pi)_\mathbbmss{a} )/\Gamma(Y\to  \psi^{'} (\pi\pi)_\mathbbmss{a}) 
\end{equation} 
The square of the matrix element weights the  three-body phase space where the two pions are the decay products of an intermediate~ scalar resonance. The Breit-Wigner ansatz we use is a rough approximation for the description of a very broad  $\sigma$ meson whereas is rather suitable for the $f_0(980)$. We find for the $Y(4350)$ the ratio $\Gamma_{\sigma}=0.3$, to be compared to the limit extracted in~\cite{cotugno}: $\mathcal{B}(Y(4350)\to J/\psi\pi^+\pi^-)/\mathcal{B}(Y(4350)\to \psi^{'}\pi^+\pi^-)<3.4\times 10^{-3} \;@ \;90\;\% {\rm C.L.}$ For the $Y_B(4660)$ the corresponding ratios due to both $\sigma$ and $f_0(980)$ modes are $\Gamma_{\sigma}=0.6$ and $\Gamma_{f_0}=1.9$, to be compared with $\mathcal{B}(Y_{B}(4660)\to J/\psi\pi^+\pi^-)/\mathcal{B}(Y_{B}(4660)\to \psi^{'}\pi^+\pi^-)<0.46 \;@ \;90\;\% {\rm C.L.}$\cite{cotugno}. Although it cannot reproduce entirely the experimental suppression of the $\Jp\pi^+\pi^-$ modes, the ansatz contained in Eq.~(\ref{matr}) is able to 
account for some enhancement of $\psi^{'}\pi\pi$ despite the phase space suppression.

\section{Decays of tetraquark states}\label{4qsectiondecays}
We give now some general considerations about the expected decay patterns of these tetraquark states. Each tetraquark wave function contains an admixture of color singlet meson pairs, and one therefore expects that hadronic decays will be dominated by a ``fall-apart'' mechanism in which the tetraquark dissociates into meson pairs, either charmed mesons or a charmonium plus light meson(s), and also radiative decays in which a light quark pair annihilates into a photon. The $\XX$ decays in all three ways.

We show in Table~\ref{tab:4qdecays} the $J^{PC}$-allowed hadronic and radiative decays of $L=0$ and $L=1$ tetraquarks and the corresponding partial waves. We also identify selection rules which arise due to the spin part of the decay amplitude, the details of which are derived in Appendix \ref{recoupling}. Entries without parentheses indicate that the fall-apart decay can proceed by a spin-conserving process, which is to say that the diquark-antidiquark spin wave function 
recouples directly into the spin wave functions of the final state mesons. Those with parentheses require the nonconservation of spin, either of the light quark or heavy-light pair (round brackets), or the heavy quark pair (square brackets); we expect the latter to be a stronger selection rule. Such arguments have been used before in the context of the $\XX$ as a $1^{++}$ tetraquark \cite{4q} or molecule \cite{Voloshin04heavy}, where the wave function in either case necessarily has the $c\bar c$ in spin-one, consistent with the dominance of the $J/\psi \pi^+\pi^-$ and $J/\psi \pi^+\pi^-\pi^0$ modes.  
\begin{table}
 \begin{tabular*}{1.0\textwidth}{@{\extracolsep{\fill}}|r|cccccc||r|ccccc|}
\hline
		&$0^{++}$
			&$1^{++}$	
				&$2^{++}$
					&$0^{-+}$	
						&$1^{-+}$
							&$2^{-+}$
								&
										&$1^{+-}$
											&$0^{--}$	
												&$1^{--}$
													&$2^{--}$	
														&$3^{--}$\\
\hline
$\eta_c\pi^0$ 	&S	&	&[D]	&	&(P)	&	&$J/\psi\pi^0$	&S,D	&(P)	&(P)	&(P,F)	&(F)	\\
$\eta_c\eta$	&	&	&	&	&(P)	&	&$J/\psi\eta$	&	&(P)	&(P)	&(P,F)	&(F)	\\
$J/\psi\rho$	&	&	&	&(P)	&(P,F)	&(P,F)	&$\eta_c\rho$	&	&[P]	&(P)	&[P,F]	&[F]	\\
$J/\psi\omega$	&	&	&	&(P)	&(P,F)	&(P,F)	&$\eta_c\omega$	&	&[P]	&(P)	&[P,F]	&[P,F]	\\
$\eta_c'\pi^0$ 	&	&	&	&	&(P)	&	&$\psi^{'}\pi^0$&	&(P)	&(P)	&(P,F)	&(F)	\\
$D\bar D$	&	&	&	&	&(P)	&	&$D\bar D$	&	&	&P	&	&(F)	\\
$D^*\bar D$	&	&	&	&P	&P	&P	&$D^*\bar D$	&	&P	&P	&P	&(F)	\\
$\eta_c\sigma$	&	&[P]	&[P]	&S	&	&D	&$J/\psi\sigma$	&(P)	&	&S,D	&D	&D	\\
$\chi_0\pi$	&	&	&	&S	&	&D	&$h_c\pi$	&	&	&S,D	&[D]	&[D]	\\
$\chi_1\pi$	&	&	&	&	&S,D	&D	&$\chi_0\gamma$	&(P)	&	&S,D	&D	&D	\\
$\chi_2\pi$	&	&	&	&D	&D	&S,D	&$\chi_1\gamma$	&	&S,D	&S,D	&S,D	&D	\\
$J/\psi\gamma$	&S,D	&S,D	&S,D	&(P)	&(P,F)	&(P,F)	&$\chi_2\gamma$	&	&	&S,D	&S,D	&S,D	\\
$\psi^{'}\gamma$&	&	&	&(P)	&(P,F)	&(P,F)	&$\eta_c\gamma$	&S,D	&[P]	&(P)	&[P,F]	&[F]	\\
$h_c\gamma$	&	&	&	&S,D	&S,D	&S,D	&$\eta_c'\gamma$&	&[P]	&(P)	&[P,F]	&[F]	\\
\hline
 \end{tabular*}
\caption{Hadronic and radiative decays of tetraquark states in given partial waves S, P, D, F. Partial waves in parentheses indicate that the decay can take place only by spin-flip, either of the light or heavy-light quark pair (round brackets) or the heavy quark pair (square brackets). In applying the rule to decays involving the $\sigma$, we assume that they proceed through an intermediate $q\bar q$ with $\an 3P0$ quantum numbers which then feeds the physical $\sigma$.}
\label{tab:4qdecays}
\end{table}

{\bf \emph{The L=0 states}}. 
The $J^+$ states present something of a problem from a phenomenological point of view, insofar as they are as light or even lighter than the corresponding $h_c$ and $\chi_c$ states, sharing the same quantum numbers and hence decay modes, and yet there is apparently no experimental evidence for their existence. Recall that among the neutral states we would expect four additional $0^{++}$ and $1^{+-}$, and two additional $1^{++}$ and $2^{++}$ states. Referring to Table \ref{tab:4qdecays}, we note that the two most numerous of the expected additional states, $0^{++}$ and $1^{+-}$, can decay in S-wave to $\eta_c\pi$ and $J/\psi\pi$ respectively, and one could argue that they simply ``fall-apart'' broadly in such a way as to be effectively unobservable. The remaining states $L=0$ cannot be dismissed in this way. The $1^{++}$ states, because of their unnatural parity, cannot decay into $\eta_c\pi$ and we thus expect them to be comparatively narrow. The corresponding $\chi_{c1}$ state has a width of less than 1 MeV; the tetraquark analogue may be broader on account of fall-apart decay into $\eta_c\pi\pi$ via the low mass tail of the broad $\sigma$, however this is forbidden in the limit of heavy quark spin conservation, since the $1^{++}$ tetraquark necessarily has the $c\bar c$ in spin 1, as discussed above. The $2^{++}$ states could decay to $\eta_c\pi$ in D-wave, but this is also forbidden by heavy quark spin conservation. In the tetraquark picture we thus expect light, narrow $1^{++}$ and $2^{++}$ states in the $\chi_c$ mass region decaying into $\eta_c\pi$ and $J/\psi\gamma$.

{\bf \emph{The L=1 states}}. 
Because of their higher mass the $J^-$ states have many more available decay modes and considerably more phase space, and it is plausible that most of them decay broadly so as to be effectively unobservable. The $\XX$ with $2^{-+}$ quantum numbers may be a unique exception. Because of its unnatural parity it cannot decay to $\eta_c\pi$, $\eta_c\eta$, or $D\bar D$; its observed decays into $J/\psi\rho$, $J/\psi\omega$ and $\ddsn$ are all P-wave with very little phase space, naturally implying a small width in accordance with the experimental data, $\Gamma=3.0^{+2.1}_{-1.7}$~MeV. Moreover, we note that the $J/\psi$ modes are suppressed by the conservation of quark spin, although because only the light quark spin needs to flip this selection rule may be badly violated. Indeed the very observation of these modes confirms that the rule is broken, but without further model-dependent assumptions it is difficult to ascertain to what extent the narrowness of the $\XX$ is due to the limited phase space and to what extent it is due to this weak selection rule. We note also that the observed $J/\psi\gamma$ and $\psi^{'}\gamma$ decays also imply the nonconservation of light quark spin: the $q\bar q$ pair can only annihilate into a photon if it has spin one, and the wave function of the $\XX$ does not contain such a component (see Appendix \ref{recoupling}). In a diquark-antidiquark model the P-wave tetraquark wave function contains a $c\bar c$ pair at large distance due to the P-wave, and thus it may be anticipated that $\chi_c\pi$ modes, particularly $\chi_{c2}\pi$, and $h_c\gamma$, should be large. We thus urge a search for these challenging modes which, if observed to be prominent, would support the tetraquark hypothesis, although as noted earlier $h_c\gamma$ should also be large in the $1\an 1D2$ case.

The pattern of allowed decays for the $0^{-+}$ states is very similar to those of the $2^{-+}$ states, largely because of their shared unnatural parity. The $\eta_c\sigma$ decay goes in S- rather than D-wave, and in a model approach \cite{4q} one expects tetraquark states to couple strongly to final states which themselves have four-quark content, as the $\sigma$ is generally accepted to have. This coupling may be so strong as to make the width of the $0^{-+}$ much larger than that of the $2^{-+}$, which may therefore make it more difficult to identify. If such a mechanism is not in place then the $0^{-+}$ states should exist and will be as narrow as the $\XX$. Indeed, they may be even narrower: due to spin-orbit splittings one expects the $0^{-+}$ to be lighter than the $2^{-+}$ \cite{Drenska:2009cd}, which would close the $\ddsn$ mode and reduce the effective phase space for $J/\psi\rho$ and $J/\psi\omega$. Such a state should, like the $\XX$, decay radiatively into $J/\psi\gamma$, and there is apparently no signal in the data \cite{AubertEtAl09evidence}.

Among the remaining states we expect the unnatural parity $0^{--}$ and $2^{--}$ to be the most narrow, although they will presumably be broader than the corresponding $0^{-+}$ and $2^{-+}$ states due to the much greater phase space available to $J/\psi\pi$ and $\Jp\eta$ compared to $\Jp\rho$ and $\Jp\omega$; if the decay amplitudes blow up rapidly as phase space opens up, these may be too unstable to be identified. If their mass is sufficient the $D^*\bar D$ mode could be prominent, while $\eta_c\rho$ and $\eta_c\omega$ are forbidden by the conservation of heavy quark spin. We expect dominant radiative decays to $\chi_{c1}\gamma$ and, for the $2^{--}$, $\chi_{c2}\gamma$. 

The remaining $1^{--}$, $3^{--}$, and $J^{PC}$-exotic $1^{-+}$ states have many decay modes available and with ample phase space, so we expect that they should be the least stable of all the possible $L=1$ tetraquark configurations. This may be helpful from a phenomenological point of view. The spectra of $1^{--}$ states in the 3800-3900 MeV mass region has been very well studied and there is apparently no evidence of an overpopulation of states, whereas in the tetraquark picture one expects a further eight neutral states alone. The phenomenology is thus only consistent with the assumption that the $1^{--}$ states, due to the large number of decay modes available and ample phase space, do not exist as stable resonances. It may be, however, that their radial excitations are more stable on account of the spatial separation of the $c\bar c$ pair, as discussed earlier in the context of the $Y(4350)$ and $Y_B(4660)$. If it is indeed the case that the radial ground states of the $L=1$ tetraquark states with $1^{--}$ quantum numbers do not exist as stable resonances, then one can infer, judging by the overall similarity of their decay patterns to those of the $1^{--}$ states, that the $3^{--}$ and $1^{-+}$ states are likewise unstable and probably do not exist. 
 
\section{Conclusions}
The BaBar result which favors $2^{-+}$ quantum numbers for the $\XX$ implies a serious revision of theoretical interpretations is required. If it is confirmed, the molecular interpretation appears to be untenable, but as we have shown the $1\an 1D2$ and tetraquark interpretations also face challenges. 

If the $\XX$ is a $1\an 1D2$ state, we would expect a much smaller prompt production cross section than that which is observed at the Tevatron. Moreover, we have confirmed the result obtained in other models that the mass is much heavier than would be expected. Our model, an extension of one previously discussed in~\cite{cotugno}, predicts with remarkable accuracy the masses of most established charmonia and bottomonia mesons, while for the $\XX$ there is a considerable discrepancy. As for the decay pattern we suggest that $h_c\gamma$ and $\eta_c\pi\pi$ ought to be prominent, and that the verification or otherwise of $\psi'\gamma $ may help clarify the situation.

We reexamined the tetraquark spectrum in the hypothesis that the $X(3872)$ is a P-wave tetraquark. The corresponding diquark mass is lower than that obtained if $X(3872)$ were a $1^{++}$ state, implying a series of S-wave states with masses comparable to P-wave charmonia. This may be difficult to reconcile with data, and, in particular, we expect narrow $1^{++}$ and $2^{++}$ states. On the other hand, among the P-wave tetraquarks in the  $3800-3900~{\rm MeV}$ region it is the $\XX$ with $2^{-+}$ quantum numbers which ought to be the most stable; many of its partner states, for which there is no experimental evidence, are probably too broad to exist as genuine resonances. On the other hand, we must emphasize that the narrowness of the $\XX$ is not a consequence of the assumed four-quark structure: any $2^{-+}$ state at 3872 MeV will be narrow because of the remarkable fact that its allowed decays are all P-wave with almost no phase space. An important consequence of the tetraquark interpretation is the reassignment of the $Y(4350)$ and $Y_B(4660)$ with respect to Ref.~\cite{cotugno}; in this picture they could be understood as the first and second radial excitations of a P-wave $[cq][\bar c \bar q]$ bound state, explaining why the $\psi^{'}\pi\pi$ decay mode dominates over the $J/\psi\pi\pi$ one despite of phase space suppression. 

\begin{appendix}
\section{Equal and Infinite heavy mass limit}
\label{app:limit}
The energy and the orbital angular momentum of a relativistic string spinning with an angular velocity $\omega$
with two masses $m_1$ and $m_2$ attached at each end can be written as a function of $T$, $m_1$, $m_2$ and $\omega$.
In the equal and infinite mass limit ($m_1=m_2=M$), we can expand Eq.~(\ref{eandorb2}) up to the second order in the parameter $\epsilon=T/M\omega$,

\be
\label{start}
\begin{cases}
{\cal E}&= 2M+\epsilon\frac{2 T}{\omega}+ \epsilon^2 M+\mathcal{O}(\epsilon^3)\sim2 M + \frac{3 T^2}{M \omega^2}\\
&\\
L&=\epsilon\frac{2 T}{\omega^2}+\mathcal{O}(\epsilon^3)\sim\frac{2 T^2}{M \omega^3}
\end{cases}
\ee

In this limit we can compute the dependence of ${\cal E}$ on the distance between the masses, using Eq.~(\ref{tension})
to eliminate $\omega$ in favor of $R=2r=r_1/2=r_2/2$:

\be
\omega^2=(1-(\omega r)^2)\frac{T}{Mr} \Rightarrow \omega=\sqrt{\frac{T}{M r + r^2 T}}
\ee

One finds
\be
\label{dist}
\begin{cases}
{\cal E}&=2 M + \frac{3 R T}{2} + \frac{3 R^2 T^2}{4 M}\\
L&=\frac{T^2}{4M} \left(\frac{2MR + R^2 T}{T}\right)^{3/2}
\end{cases}
\ee
which reduce to Eqs. (\ref{Elimit}) and (\ref{Llimit}) for $M>>TR$. From the first of these two relations it is evident that this relativistic string model well describes the confinement, the energy being a growing function of the 
distance between the quark and the antiquark inside the meson.
The second of the relations in~(\ref{dist}) allows to write the distance $R$ between the quarks in terms of the orbital angular momentum, the mass, and the tension:
\be
\label{erremio}
R=-\frac{M}{T} + \frac{\sqrt{M^2 T^{2/3} +  (4LMT^2)^{2/3} }}{T^{4/3}}
\ee
which reduces to Eq. (\ref{Rlimit}) for $M>>TR$.

\section{Relativistic Corrections}
\label{app:rel}
The spin-orbit term contains the derivative of the energy $E$ with respect to the distance $R$. 
Using the first relation in Eq.~(\ref{dist}) gives
\be
\label{derivative}
\frac{\partial E}{\partial R}=\frac{3 T (M + R T)}{2 M}
\ee
The spin orbit term can thus be rewritten in terms of $T$, $M$ and $L$ plugging the expressions found for $R$ in~(\ref{erremio}) inside Eq.~(\ref{derivative}). For the tensor term the magnetic field generated from a magnetic moment $\pmb{\mu_1}$ is
\be
\pmb{H}_1=\frac{3\,\pmb{n}\,\left(\pmb{\mu}_1\cdot \pmb{n}\right)-\pmb{\mu}_1}{R^3}
\ee
The interaction energy is
\be
\label{tensor}
M_{\mu\mu}=-\pmb{\mu}_2\cdot \pmb{H}_1-\pmb{\mu}_1\cdot \pmb{H}_2=-\frac{3\left(\pmb{\mu}_2\cdot\pmb{n}\right)\,\left(\pmb{\mu}_1\cdot \pmb{n}\right)-\pmb{\mu}_2\cdot\pmb{\mu}_1}{R^3}+(1\leftrightarrow 2)
\ee
where $\pmb{n}$ is the unit vector in the direction between the two heavy quarks. With
\begin{equation}
\label{defeff}
\pmb{\mu}_i=\frac{ge_i}{2m_i}\pmb{S}_i
\end{equation}
\noindent and
\begin{equation}
\pmb{S}^2-3\left(\pmb{S}\cdot\pmb{n}\right)^2=2\left[\pmb{S}_1\cdot\pmb{S}_2-3\left(\pmb{S}_1\cdot\pmb{n}\right)\left(\pmb{S}_2\cdot\pmb{n}\right)\right]
\end{equation}
one finds
\be
M_{\mu\mu}=B\frac{1}{R^3}\left[\pmb{S}^2-3\left(\pmb{S}\cdot\pmb{n}\right)^2\right]\\
\ee
\noindent Again, using the expression~(\ref{erremio}) we can write the tensor interaction as a function of $T$, $M$ and $L$.

\section{Spin recoupling matrix elements}\label{recoupling}
The diquark-antidiquark spin wavefunctions in the $(S_{cq}\times S_{\bar c \bar q})_{S}$ basis can be rewritten in the bases $(S_{c\bar q}\times S_{q \bar c})_{S}$ and $(S_{c\bar c}\times S_{q \bar q})_{S}$ of the open- and closed-flavor final states, respectively. The matrix elements $\bk<(S_{c\bar c}\times S_{q\bar q})_S|(S_{cq}\times S_{\bar c\bar q})_S>$ for closed flavor decay are
\be
\begin{array}{rcc}
			&\k|(0\times 0)_0>	&\k|(1\times 1)_0>	\\
\b<(0\times 0)_0|	&1/2			&\sqrt{3}/2		\\
\b<(1\times 1)_0|	&\sqrt{3}/2		&-1/2			\\
\end{array}
\ee
\be
\begin{array}{rccc}
			&\k|(1\times 1)_1>	&\sqrt{1/2}\left(\k|(1\times 0)_1>+\k|(0\times 1)_1>\right)
									&\sqrt{1/2}\left(\k|(1\times 0)_1>-\k|(0\times 1)_1>\right)\\
\b<(1\times 1)_1|	&0			&1			&0			\\
\b<(1\times 0)_1|	&\sqrt{1/2}		&0			&\sqrt{1/2}		\\
\b<(0\times 1)_1|	&\sqrt{1/2}		&0			&-\sqrt{1/2}		\\
\end{array}
\ee
\be
\begin{array}{rc}
			&\k|(1\times 1)_2>			\\
\b<(1\times 1)_2|	&1					
\end{array}
\ee

The corresponding matrix elements $\bk<(S_{c\bar q}\times S_{q\bar c})_S|(S_{cq}\times S_{\bar c\bar q})_S>$ for open flavor decay are:
\be
\begin{array}{rcc}
			&\k|(0\times 0)_0>	&\k|(1\times 1)_0>	\\
\b<(0\times 0)_0|	&-1/2			&\sqrt{3}/2		\\
\b<(1\times 1)_0|	&-\sqrt{3}/2		&-1/2			\\
\end{array}
\ee
\be
\begin{array}{rccc}
			&\k|(1\times 1)_1>	&\sqrt{1/2}\left(\k|(1\times 0)_1>+\k|(0\times 1)_1>\right)
									&\sqrt{1/2}\left(\k|(1\times 0)_1>-\k|(0\times 1)_1>\right)\\
\b<(1\times 1)_1|	&0			&0			&-1			\\
\b<(1\times 0)_1|	&\sqrt{1/2}		&-\sqrt{1/2}		&0			\\
\b<(0\times 1)_1|	&\sqrt{1/2}		&\sqrt{1/2}		&0		\\
\end{array}
\ee
\be
\begin{array}{rc}
			&\k|(1\times 1)_2>			\\
\b<(1\times 1)_2|	&1					
\end{array}
\ee

In deriving the selection rules of Table \ref{tab:4qdecays}, we bear in mind that the physical states for each $J^{PC}$ can contain any admixture of the various possible spin wave functions in Table \ref{tab:tetra}, and we thus consider only those selection rules which apply to all possible admixtures.

\end{appendix}

\bibliography{paper}

\end{document}